\documentclass[aps,prl,twocolumn,groupedaddress,subfigure,showpacs]{revtex4}
  \usepackage[latin1]{inputenc}

  \usepackage{epsfig}
  \usepackage{graphicx}
  \usepackage{amssymb}

\newcommand{\delete}[1]{{}}

\begin{document}

%\title{ Spectra of Stable Single Bubble Sonoluminescence with no Noble Gasses Present}

\title{ Spectra of Stable Non-Noble Gas Single Bubble Sonoluminescence}

\author{Mogens T. Levinsen}
\author{Jeppe Seidelin Dam}
\email{levinsen@nbi.dk}

\affiliation{Complexity Lab, Niels Bohr Institute,
                 Blegdamsvej 17, DK-2100 Copenhagen {\O}, Denmark}

\date{\today}

\begin{abstract}
A commonly accepted view is that stable Single Bubble Sonoluminescence (SBSL)
can only be achieved in the presence of a noble gas or hydrogen. In air-seeded 
bubbles, the content of diatomic gasses is burned off to leave the small 
amount of argon needed to sustain stable operation. Here we report that long 
term stable SBSL can be sustained with only nitrogen, oxygen, or 
nitrogen/oxygen mixtures being present. Compared to that of a stable argon 
bubble, the emission is much weaker and the spectrum looks much colder. 
Oscillating states as well as recycling states are also observed. An 
intriguing saturation effect seems connected with the presence of water 
vapor in the bubble. 

\end{abstract}


\maketitle

Recently we reported a second type of recycling mode in Single Bubble 
Sonoluminescence (SBSL) \cite{dam-type2}. This occurs for air-seeded bubbles 
with dissolved gas content higher than approximately 20\% of the saturation 
level, when driven past their shape instability limit. In contrast 
to the usual recycling mode denoted type 1, which reaches the same maximum 
intensity level as that of the stable bubble,
the intensity in the new type 2 mode is down by approximately 
a factor of 20. Furthermore, spectra of the emission from a single bubble in 
the type 2 mode look like the bubble is much colder than an argon dominated 
bubble, thus providing a link to multi bubble sonoluminescence (MBSL). 

The dissociation hypothesis \cite{lohse} was invented to explain the 
stability of SBSL air-seeded bubbles at dissolution levels, where the
bubble growth elsewise would not be limited. This hypothesis, since
verified in many experiments, suggests that the bubbles burn off 
their content of diatomic gasses leaving behind the small amount of argon 
needed to sustain stable operation. The chemical reaction 
products, readily dissolvable, diffuse out of the bubble during the 
expansion and following contraction. Our interpretation of the type 2 mode 
follows this picture. The bubble is driven so hard that after the splitting 
off of a microbubble as is the usual explanation for the type 1 event, it 
sucks in air too fast to burn off the diatomic molecules.  Thus it 
grows to the instability limit while still being predominantly an air bubble. 
The role of the water vapor could not be established.   

In order to check this hypothesis we prepared a cell with  
gas mixtures involving no noble gasses (nor hydrogen). Gasses used are
either pure nitrogen, oxygen, or a 4:1 nitrogen-oxygen mixture to simulate 
air without argon. The noble gas content in all cases are
below 1ppm. 

As we had expected, SBSL in the type 2 recycling state is possible and at 
the same level of emission. The spectrum also looks very similar to 
that obtained from air-seeded bubbles. Thus our hypothesis regarding the 
type 2 recycling seems verified.
 
However, to our surprise, stable SBSL was also possible. This has to the 
best of our knowledge not been reported before.  The stable radiation was 
obtained for extended periods of up to 45 minutes for nitrogen bubbles. The 
fluctuations and long term drift of the relative phase of the flashes 
were 100 ns or less with short term fluctuations within 40 ns and no preferred 
direction of drift.
At higher levels of gas pressure an oscillating state is encountered at the 
low end of the emission range. Such an oscillating state has previously been  
reported  for nitrogen bubbles \cite{Hiller}. 
These authors also reported on irregular peaking emission. This phenomenon 
we find to be associated with small amounts of argon being allowed to diffuse 
into the cell. 

The spectrum of the stable phase very much coincides with that of the 
type 2 case for air seeded bubbles with total output being similar
in the case of nitrogen and the nitrogen/oxygen mixture but an order of 
magnitude lower for pure oxygen which also seems to be colder. 
Thus this experiment is further verification of the dissociation
theory and fits the picture of the strong and hotter emission from an 
air-seeded bubble being connected with argon accumulation. 
To obtain stable emission from non-noble gasses, however, must call 
for a delicate balance between the diffusion into the bubble and 
the burning of diatomic gasses. Presumably this leaves the bubble 
much more susceptible to environmental factors like microstreaming, 
nearby particles, and chemical balance. All factors, that are expected 
to affect the long term stability. In view of this, while the levels of 
fluctuations and long term drift are higher than would be expected for 
e.g. air or argon seeded bubbles at high levels of  emission which can 
be stable at ns levels, the changes are still surprisingly small.  

The cell used is described in more detail in ref.~\cite{levinsen}, but 
is essentially a 5 or 6 cm high and 6 cm diameter cylindrical quartz cell 
with metal caps at both ends so it can be sealed using a pressure relief bag. 
Piezoelectric transducers are mounted on one or both caps for the drive, 
which for most experiments reported (6 cm cell), is at a  frequency of 
22140 Hz. A notable difference is the use of a heater to avoid contamination 
with hydrogen.

The procedure finally adopted to prepare the water is as follows. The water 
is subjected to alternating degassing at water vapor pressure and flushing 
with the final gas to be used for several hours in 15 minutes intervals. 
Under this whole process the water is violently stirred by a magnetic stirrer. 
After a final 15 min. flushing with the final gas composition and 
concentration, the water is  transferred in a closed system to the cell and 
cooled to the operating temperature ($~ 9 ^\circ C$) for several hours before 
commencing measurements.

As described in ref.~\cite{dam-type2} a crude spectrum is obtained by 
placing 7  photomultiplier tubes (PMT's) around the cell with narrow band 
(10 nm) optical filters in front. The signals are amplified and 
collected by a computer with an extra PMT giving the timing signal for 
the sampling. This method was chosen primarily because it allows us to get 
spectra of weak and shortlived states but also since it allows us to 
get flash by flash control of the averaging time of the spectra. Thus we 
have precise knowledge of the time of collection so the radiative state of 
the bubble can be established. The timing of the flash is also recorded with 
a resolution of 50 ns with averaging providing a better resolution for 
slower fluctuations. For extremely weak signals (oxygen bubbles), 
however, a different procedure, to be described later, was used.

\begin{figure}[t]
\centering
\mbox{}
\vspace{-0.4cm} 
\mbox{\epsfig{figure=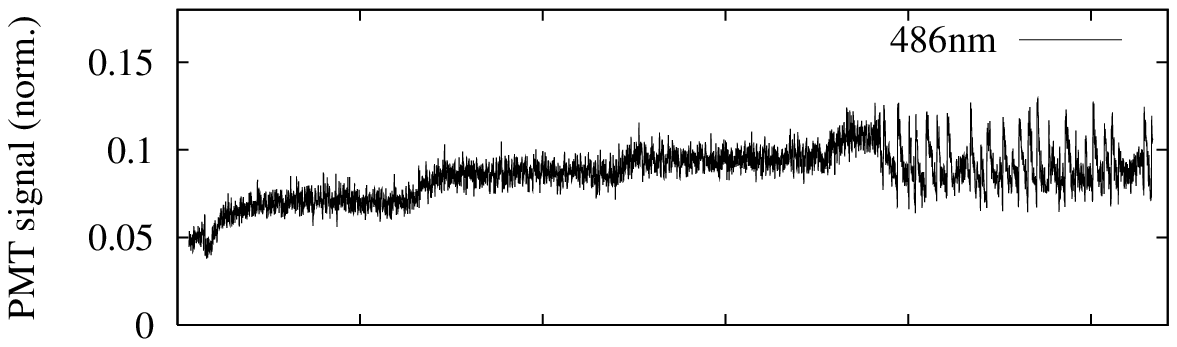,width=8.5cm}}
\vspace{-0.4cm}
\mbox{\epsfig{figure=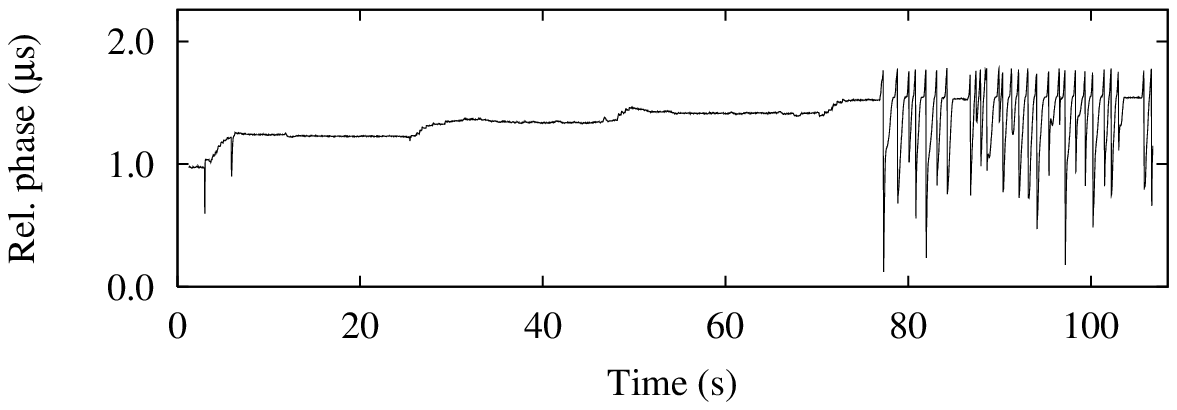,width=8.5cm}}
\caption{Timeseries showing a progression for stepwise increase of 
sound pressure through stable states into type 2 
behavior. Gas content 880 mbar nitrogen/oxygen 4:1 mixture as prepared at room 
temperature. Measurement performed at 9$^{\circ}$C. Upper trace emission at 
486 nm $\pm$ 5nm. Lower trace the relative phase. Average over 1000 flashes.}
\label{const_timeseries}
\end{figure}

The stable states were in the case of the nitrogen/oxygen mixture found in 
the whole interval investigated from dissolved gas levels of 180  to 900 mbar. 
For nitrogen the range of degassing where a stable state was observed 
is 100 to 320 mbar, with the intensity level falling below our limit of 
detection at the low end. However, type 2 emission was observed up to 900 mbar.
At low levels of dissolved gas, the interval of stability is quite narrow with 
the range of stability seemingly being wider for pure nitrogen than for the 
nitrogen/oxygen mixture. Near ambient pressure the intervals open up and 
we are able to measure the change in spectrum versus amplitude of the drive. 
Apart from type 2 (only recycling state present), we also see the 
oscillating state observed by Hiller et al \cite{Hiller}. 
These states look very much like those recently observed for
air-seeded bubble at very low drives by Thomas et al \cite{thomas}. 
This suggests that these low intensity states from air-seeded bubbles 
have the same origin of being mainly nitrogen/oxygen bubbles. 

\begin{figure}[t]
\centering 
\mbox{\epsfig{figure=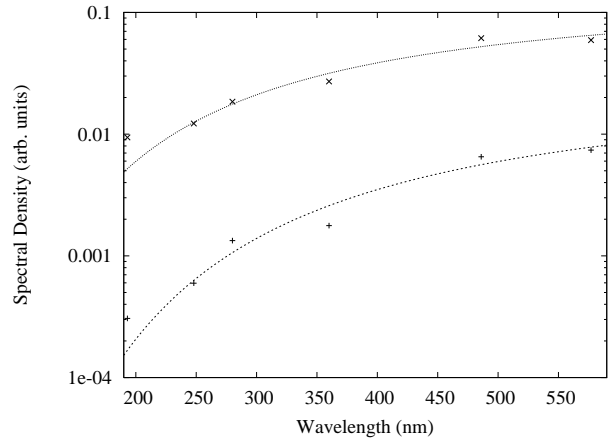,width=8.5cm}}
%\vspace{1cm}
\caption{Spectra of constant emission $(\times)$  dissolved gas level 220 nitrogen, $(+)$ 220 mbar oxygen normalized with emission from stable air bubble. The curves are fitted black body radiation normalized within a constant with black body radiation at 14000 K. Upper curve T $=$ 8000 K, lower curve 6580 K. Drive frequency 23200 Hz} 
\label{amp_var_stable}
\end{figure}

With gas levels of more than approximately 200 mbar, we encounter the type 2
recycling. This has essentially the same time scale of recycling 
as the air-seeded bubbles, which strongly supports our explanation
of type 2 recycling having its origin in incomplete burn-off. 
Also it indicates the size $R_0$ of the bubbles at breaking
point to be of the same scale of approximately 6 $\mu$m.
This is corroborated by observation of an anisotropic 10\% period 
doubling in the measurements of the 280 mbar nitrogen bubble just before 
and at the level of type 2 recycling. 

In Fig.~\ref{const_timeseries} we display a 100 s long timeseries
obtained for a single bubble in water prepared  at a dissolved gas 
level of 880 mbar (nitrogen/oxygen (4:1) mixture) at room
temperature and cooled to 9$^{\circ}$C. The amplitude of the drive is
stepped up rapidly at 4 distinct short time intervals with stable regions of
emission in between, before finally type 2 recycling sets in. Note the 
overlap in regimes, as the bubble several times relapses into 
the quiet state, although the sound pressure is kept constant in this 
last part of the sequence. Since the emission is weak and narrow band 
optical filters are employed, the average number of photons recorded per 
flash is of the order of 1. Our claim of stability is therefore based 
on the absence of oscillations or peaking and the stability of the 
relative phase. 

In Fig.~\ref{amp_var_stable} we show spectra taken for stably emitting 
nitrogen bubbles (dissolved gas level of 220 mbar) and oxygen bubbles 
(220 mbar). Especially the oxygen bubble is so weak that it can hardly be 
discerned by the unaided eye even after long time adjustment to complete 
darkness. These spectra are therefore obtained by freerunning the data 
acquisition cards at 5 MHz alternatingly for a prolonged period. 
Using the information from the phaselocked sawtooth signal fed to both cards 
one can then obtain averaged flash intensities from the remaining 6 channels
with a highly improved signal/noise ratio.
 
The spectra have been normalized by the spectrum of 
an air bubble driven at a sufficiently low level to ensure that the 
spectrometer is not overloaded. The latter spectrum has in a separate 
experiment been shown to fit well to  14000 K blackbody radiation in the 
regime of 300 nm - 700 nm but dropping below this in the extreme ultraviolet 
range (VUV) (see e.g. \cite{bb}). The spectra can be well 
fitted to black body radiation with temperatures 8000 K and 6580 for nitrogen 
respectively oxygen with the slight overshoot in the VUV range caused by the 
corresponding drop in the normalizing spectrum. By comparison a stable 
weakly radiating air-seeded bubble still displays a spectrum like that of 
the reference bubble apart from a scaling factor. The temperatures obtained 
are remarkable close to the expected dissociation temperatures for the gasses 
involved.

\begin{figure}[t]
\centering 
\mbox{\epsfig{figure=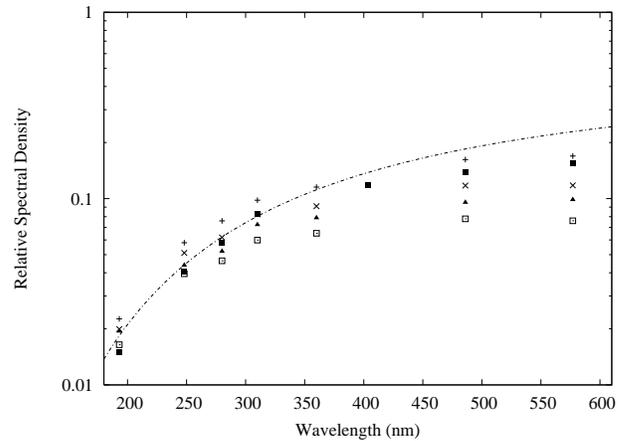,width=8.5cm}}
%\vspace{1cm}
\caption{Spectrum of the type 2 recycling state at the end of Fig. 1. From bottom to top the drive amplitude is {\small $(\boxdot)$} 130.0 mV, $(\blacktriangle)$ 130.5 mV, $(\times)$ 132.0 mV, (+) 133.4 mV. As seen by comparison to Fig.~2 there is an overlap with the constant emission state.  For comparison we have included the spectrum of a type 2 state measured for a air-seeded bubble at the same level of dissolution {\small $(\blacksquare)$}. The curve is black body radiation at 8000 K normalized with black body radiation at 14000 K for comparison.}
\label{type2}
\end{figure}

The change in the spectrum when increasing the amplitude in the type 2 
case displayed in Fig.~1 is shown in Fig.~\ref{type2}. The spectrum starts 
out looking very much like those obtained for the stable case but gradually 
change to look like a blackbody 
spectrum. For comparison we have included the spectrum of a type 2 state 
measured for a bubble where the dissolved gas is air at the same level 
of dissolution. In these figures the spectra are obtained as a long term 
average over many periods of the recycling. As seen the spectra look 
very similar giving further credence to the assumption 
that the mechanism behind is the same.   

The apparent temperatures are what would be expected from bubbles that are 
inefficiently trying to burn the diatomic gasses. However, an interpretation 
in terms of actual temperatures, though tempting, is questionable, as the 
mechanism behind the emission is still unclear. 

In the following we shall concentrate on the nitrogen bubbles to keep the 
chemical picture as simple as possible.
At the low end of the nitrogen concentration we only see the stable state. At 
higher nitrogen levels we see the oscillating state akin to that observed by 
Thomas et al \cite{thomas}. The main feature of this state is an oscillation 
seen at all wavelengths so that the color temperature stays nearly constant,
and with amplitude in phase with a corresponding oscillation in radius. 
Unfortunately we are not able to calibrate our system well enough to tell 
the exact relation.

If even small amounts of argon is allowed to diffuse into the cell, the 
behavior changes dramatically. In what follows we show the result of a 
simple analysis on some examples of states observed for a nitrogen seeded 
bubble at 28 \% saturation contaminated with argon.

Since we are not able to follow the bubble movements simultaneously with 
obtaining the spectrum, we resort to using the relative phase for calculating 
the bubble size \cite{gabor1}. The model is based on the Rayleigh-Plesset 
equations taking into account water vapor but disregarding the chemical 
processes. A problem is that 
for obvious reasons we are not able to use the dissociation theory to 
calculate the actual composition of the gas inside the bubble as we 
clearly deal with incomplete burn-off.

The calibration can thus only be based on two fixpoints, the occurrence 
of period doubling, and the onset of shape 
instability leading to type 2 recycling.  This fixes the bubble size to 
approximately  6-7 $\mu$m at the onset of type 2 behavior. The absolute 
pressure $P_a$ is more difficult to assess. However, a knowledge of the 
cell gained from measurements on air seeded bubles leads us to adopt the value
1.5 bar for the corresponding pressure. Assuming a constant average 
concentration of nitrogen in the bubble, a model calculation leads to  
c $\approx$ 0.001 compared to the degassing level of c $\approx$ 0.3.
From the diffusion equation we find that the bubble must burn approximately
$10^5$ molecules of nitrogen per flash which is the same order of magnitude 
as found for an air-seeded bubble \cite{suslick}. 

\begin{figure}[t]
\centering 
\mbox{\epsfig{figure=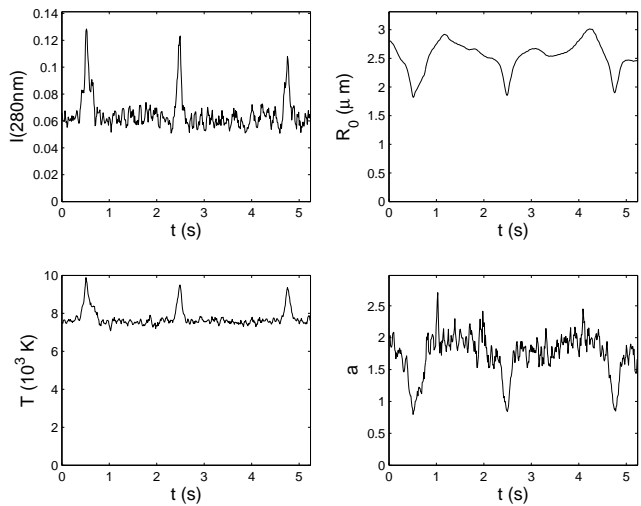,width=8.5cm}}
%\mbox{\epsfig{figure=Figx-1540.eps,width=8.5cm}}
%\mbox{\epsfig{figure=Figx-1513a.eps,width=8.5cm}}
%\mbox{\epsfig{figure=Figx-1491.eps,width=8.5cm}}
%\mbox{\epsfig{figure=Figx-1444.eps,width=8.5cm}}
%\vspace{1cm}
\caption{Timeseries obtained for a nitrogen bubble prepared with 280 mbar gas 
content showing a peaking state. The sound pressure converted from the drive 
voltage as described in the text is approximately 1.42 bar, Also shown are the 
fits for the ambient radius $R_0$, apparent temperature T, and prefactor a.}
\label{analysis}
\end{figure}

The spectral temperature is determined by fits to the blackbody spectrum which 
for the nitrogen spectrum fits quite well. For the oscillating state
at drive levels below that of the stable state and a transient 
state into the stable state, the apparent temperature also oscillates or peaks.
For the transient state (results shown in Fig.~\ref{analysis}), for most of 
the time the light emission is quite low compared to the stable emission. 
In the peaks, however,  the emission goes up at all wavelengths but most in 
the ultraviolent and actually overshoots the stable emission. At the same 
time the bubble is decreasing in size, the inference being that the bubble 
is trying to burn off nitrogen and turn into an argon dominated bubble.
Due to the normalization with the known spectrum of an argon bubble, the
prefactor is proportional to a product of emitting surface area 
and duration of flash. Exact knowledge of these factors would be 
necessary for a complete analysis. Unfortunately, both parameters would 
be very hard to measure due to the short timescale involved. It is clear 
though, that the lower the drive and the smaller 
the bubble, the higher the apparent temperature. This trend, also found 
for air-seeded bubbles, could be connected with the presence of water vapor 
\cite{toegel}. 

\begin{figure}[t]
\centering 
\mbox{\epsfig{figure=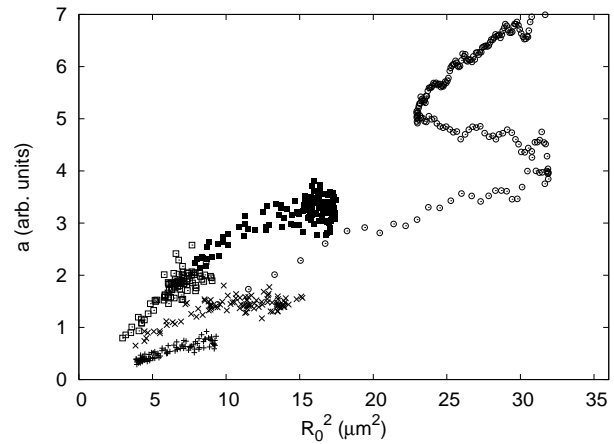,width=8.5cm}}
%\vspace{1cm}
\caption{The prefactors as function of ${R_0}^2$. $P_a$ 1.36 ({\small +}), 1.39 ({\small x}), 1.42 ({\small $\boxdot$}), 1.45 bar ({\small $\blacksquare$}). Air-seeded bubble in type 1 mode ({\small $\odot$}).} 
\label{prefactor}
\end{figure}

Although it is difficult to get quantitative information from this analysis, 
some interesting information arise from assuming the bubble radius at the 
time of emission to be proportional to R$_0$, with a proportionality factor 
that is relatively independent of drive amplitude, and plotting the prefactor 
as function of the radius $R_0$ squared. 
In Fig.~\ref{prefactor} we present this kind of plot for an oscillating 
state and a peaking state below the stable state, the transient state, and a 
peaking state above the stable state, together with an analysis for a type 1 
state (air-seeded) at a similar level of degassing for comparison. As seen, 
in the case of pure nitrogen, the prefactors at small ambient radii are 
proportional to ${R_0}^2$, but at higher values of the ambient radius  
saturation sets in. Naively one could therefore interpret the slopes to be 
proportional to the flash duration. A tentative explanation for the saturation
could be that the actual volume of emission is smaller \cite{dam-size,1MHz} 
than the actual bubble size. As the apparent temperatures do not show much 
variation, this interpretation is in fact relatively robust independent of 
whether volume or surface emission is assumed. In fact assuming volume 
emission would aggravate the problem of reconciliating the changes 
in emission with the changes in bubble size. We wish to stress 
that while the absolute values of $R_0$ and $P_a$ are naturally questionable, 
the functional dependences displayed in Figs. 4 \& 5 are robust.

The result for the recycling state of an air-seeded bubble is quite different.
The recycling is of type 1 but having the hump that eventually results in 
type 2 recycling at higher levels of applied sound pressure. For small 
bubble radius, where the bubble according to our analysis of recycling 
states is mostly an air bubble we find the same linear behavior as for the 
nitrogen bubble ending with a trend towards saturation before
the shrinking associated with the burn-off takes over.
Finally we observe a new linear increase associated with the continued 
argon intake. Stable states are all located to the right of the transient 
curves.

To conclude we have shown the possibility of obtaining stable SBSL using
non-noble gasses with total removement of noble gasses as a critical condition
for stability. The experiments confirm the interpretation of the newly found
type 2 recycling mode as originating in incomplete burn-off of diatomic gasses.
 The presence of stable emission is only permitted if the different gasses are 
prohibited from accumulating in the bubble. In the case of nitrogen, the only 
chemical processes possible involve water vapor. This 
raises the intriguing question of whether the stability could be caused by 
hydrogen actually accumulating in the bubble.  Such an accumulation 
is, however, not possible for the stable oxygen seeded bubble where the 
only possible reaction products are H$_2$O$_2$ and O$_3$. It is worth 
noting that the black body fits give $\approx$ 8000 K for nitrogen, and 
$\approx$ 6600 K for oxygen which could be related to the known dissociation 
temperatures for these diatomic molecules. Finally, gas depletion 
in a boundary layer around the bubble might very well be an 
important factor. Furthermore, both the apparent higher temperatures for 
smaller bubbles and the saturation points to the amount of water vapor being 
an important factor. Obviously an extension of the dissociation hypothesis 
also for air-seeded bubbles at low drive and/or low content argon is needed.  

We thank T. Matula and K. Musztacs for stimulating discussions. The work is 
supported by the Danish National Science Foundation.


\begin{thebibliography}{99}
\bibitem{bb}Note that this normalization should not be interpreted as 
         the emission from the air-seeded bubble necessarily being 
         blackbody radiation.
\end{thebibliography}
\end{document}